# Spectral classification and HR diagram for symbiotic stars in the galactic bulge

G. A. Medina Tanco and J. E. Steiner

Instituto Astronômico e Geofísico - USP
Av. Miguel Stefano 4200 - CEP 04301-900- São Paulo - SP - Brazil

**Abstract**

A sample of 45 symbiotic stars observed in the direction of the galactic bulge are classified in spectral type using TiO λ 6180, 7100 Å and VO λ 7865 Å absorption feature indices and visual comparison. The effective temperatures obtained are used together with K data from the literature to estimate the bolometric luminosity of the giant components if located at *d=8.5* kpc. The ($T_{eff}$, $L/L_\odot$) pairs are plotted on the HR diagram together with evolutionary tracks for metal rich intermediate mass stars. The mass of the giant components seem to be scattered in the interval ( *~1.5 - ~4 $M_\odot$* ). The giant components of D-type symbiotics appear as if they were more evolved objects than the corresponding giants in S-type systems. Nevertheless, it is unlikely that they form an evolutionary sequence. The results point also to the restriction of the symbiotic phenomenon to the thermally pulsating phase of the AGB, strengthening an accelerated mass loss scenario for the latest evolutionary stages of intermediate mass stars.

# I. Introduction

The current view of symbiotic stars is that they are binary systems formed by a red giant and a hot source ionizing the wind of the cold companion (e.g., Michalitsinos, Kafatos and Hobbs 1980; Altamore et al. 1981, Stencel et al. 1982; Kenyon and Webbink 1984; Kenyon 1986, etc.).

The resulting spectrum is the sum of several components, mainly: the hot star (and probably an accretion disk in some cases), the ionized nebula, the cold giant and the dust shell surrounding it in many cases. However, at visual and near infrared wavelengths, the spectrum is dominated mainly by the emission lines originated in the nebula and the giant's continuum characterized by strong TiO absorption features. Hence, this is the best spectral region to study the cold component, which is usually an M-type giant star (Allen 1980; Kenyon and Fernandez-Castro 1987; Kenyon 1988; Kenyon Fernandez-Castro and Stencel 1988, Schulte-Ladbeck 1988). The precise determination of the evolutionary stage of symbiotic giants is important from the point of view of stellar evolution, since they offer a unique opportunity to study the physical characteristics of the wind of an AGB star (one of the obscure points in evolutionary models) due to the illumination of the ionizing source.

Several efforts have been done in this direction, among which we can mention the work of Whitelock (1987), Schild (1989), and Seal (1990). The first two of these papers center on D-type symbiotics, and prove that the cold companion in this subclass is more likely an extreme Mira star, probably in the transition to short period OH/IR sources. The work of Seal (1990), in particular, presents an HR diagram of both, S and D-type symbiotic stars. He claims that most S-type stars match the *$3 M_\odot$* evolutionary track of metal rich giants, while D-type symbiotics are too cold to match any evolutionary track.

To minimize distance determination problems, and to have complete control over effective temperature estimation, we made spectroscopic observations of symbiotic stars (selected from Kenyon (1986) and Allen's (1984) catalogue) in the direction of the galactic bulge (< *20°* from the galactic center) and made our own spectral classification using TiO and VO absorption indices (e.g., O'Connell 1973; Kenyon 1986; Kenyon and Fernandez-Castro 1987), complemented with visual comparison with standard spectra (e.g., Jacoby, Hunter and Christian 1984; Jaschek and Jaschek 1987, Schulte-Ladbeck 1988). The results of this classification, together with the assumption of black body emission and *d=8.5* kpc, are used to locate the objects in an HR diagram, where they can be faced with evolutionary tracks of (metal rich) intermediate mass stars. For a subset of the sample, $(J-K)_0$ data compiled from the literature is used to estimate $BC_K$ and to calculate the actual luminosity of the giant components. Both, HR and magnitude-color ($K_0$, $(J-K)_0$) diagrams are presented in this case. The latter also includes isolated M and Mira giant stars of the galactic bulge which can thus be easily compared to the cold giants contained in symbiotic stars.

Spectral classification is treated in Section II, while Section III is devoted to evolutionary analysis and the closing conclusions are given in Section IV.

## II. Observations and Spectral Classification

The spectroscopic measurements were obtained with the Cassegrain spectrograph of the 1.6 m telescope of the LNA at Brazopolis, Brazil. A UV coated GEC CCD with 22 μ pixel size was used. A *300 l/mm* grating was employed resulting in 10 Å resolution. The slit width was 3.7 arcsec and typical seeing was 1-1.5 arcsec. Spectrophotometric standard stars were observed every night and the data were reduced with IRAF[1] on the Sun workstations at the Instituto Astronomico e Geofisico/USP.

Optical and near infrared spectroscopic data were obtained for 45 symbiotic stars in the direction of the galactic bulge ( $(l^2+b^2)^{1/2} \leq 20°$) in the period May-July, 1993. Table I lists the stars and Figures A.1 - A.45 show the corresponding spectra.

Our classification method is a combination of absorption band indices and visual comparison with standard spectra (e.g., Jacoby, Hunter and Christian 1984; Jaschek and Jaschek 1987, Schulte-Ladbeck 1988) in which the latter ingredient has the last word.

The absorption band indices measure the depth of the band at a central wavelength with respect to the interpolated continuum at the same wavelength (see, O'Connell 1973):

$$I(\lambda_0, \lambda_1, \lambda_2, \Delta\lambda) = -2.5 \log \left\{ \frac{F(\lambda)}{F(\lambda_1) + \left(\frac{\lambda_0 - \lambda_1}{\lambda_2 - \lambda_1}\right)[F(\lambda_2) - F(\lambda_1)]} \right\} \quad (1)$$

where $\lambda_0$ is the feature wavelength and $\lambda_1$ and $\lambda_2$ are the sidebands. $\Delta\lambda$ is the feature bandwidth, and is taken as $\Delta\lambda = 30$ Å in the infrared as a compromise between resolution and signal-to-noise ratio. O'Connell (1973) gives several features at optical and infrared wavelengths suitable for spectral classification, from which Kenyon and Fernandez-Castro (1987) adopted the TiO λ 6180, 7100 Å features, together with the VO band at 7865 Å for their spectral classification of symbiotic stars. For this purpose, they defined the indices:

$$[TiO]_1 = I\ (6180, 6125, 6370, 30) \quad (2.a)$$

$$[TiO]_{2a} = I\ (7100, 7025, 7400, 30) \quad (2.b)$$

$$[VO] = I\ (7865, 7400, 8050, 30) \quad (2.c)$$

---

[1] **IRAF** is distributed by National Optical Astronomy Observatories, which is operated by the Association of Universities for Research in Astronomy, Inc. (AURA) under cooperative agreement with the National science Foundation.

which they calibrated as a function of the spectral type (ST) for K-M stars:

$$ST = 7.75 + 9.31 \times [TiO]_1 \tag{3.a}$$

$$ST = 7.83 + 10.37 \times [TiO]_{2a} - 3.28 \times [TiO]_{2a}^2 \tag{3.b}$$

$$ST = 6.74 + 19.3 \times [VO] - 17.22 \times [VO]^2 \tag{3.c}$$

where $ST=0$ stands for K0, $ST=1$ for K1, $ST=6$ for M0, and so on. On the other hand, Kenyon (1986) defines $[TiO]_2$ in a slightly different way, namely:

$$[TiO]_{2b} = I\,(7100, 7050, 7400, 20) \tag{4}$$

and a fit to his Figure 3.13 of $[TiO]_{2b}$ vs. spectral type gives:

$$ST \cong \left[ 628 \times \ln(1 + [TiO]_{2b}) \right]^{0.4} \tag{5}$$

where, again, $ST=0$ stands for K0. We applied both equation (3.a-.c) and (5) in classifying our spectra.

Although an automatic classification was attempted using these indices, it proved to be unreliable in many cases. For several objects, in fact, either the quality of the observation or the intrinsic characteristics of the spectrum (e.g., strong emission either filling the absorption band or contaminating the sidebands) made it impossible even to define the corresponding indices or, at least, to believe the results. Furthermore, despite the fact that all the four indices gave almost the same average results for the whole sample, case by case differences were important most of the times. This can be seen clearly in Table II, where averaged spectral types evaluated using each one of the indices are given[2]. In Table II, $\langle ST \rangle$ is the mean spectral type for the complete sample of exposures, and

$$\sigma_{i,j}^2 = \frac{1}{n} \times \sum_{l=1}^{n} \left\{ [ST]_l(i) - [ST]_l(j) \right\}^2 \tag{6}$$

---

[2] The different indices were applied to every single spectrum observed (including many exposures of the same object in the same and/or different nights). Consequently, these averaged values have no relationship with the actual value of the final classification given in the paper. These numbers correspond only to a test over the whole rough data.

where $[ST]_l(i)$ is the spectral type associated with spectrum $l$ by using the index $i$. Therefore, despite the fact that the four indices give essentially the same averaged spectral type for the sample, differences in spectral type for a single object can range from 1.5 to 3 points on the average in going from one index to another. Due to these problems, we decided for a two step process. First, the spectra were automatically classified by using relations (3.a-.c) and (5) and, second, this first approximation was fine tuned by visual analysis. As an example, the results obtained from $[TiO]_{2b}$ (equation (5) - whenever it was meaningful to evaluate this index), and our final guess to the actual spectral types, are given in columns 4 and 5 of Table I respectively. The classification of Kenyon (1986) is also included in column 3 for comparison. In Figure 1 we plot the spectral classification obtained from $[TiO]_{2b}$, equation (5), vs. the final (visual) classification of this work for those stars for which the $[TiO]_{2b}$ index could be defined. It can be seen that the final differences are not too large, on the average, and that a good correlation ($\rho = 0.94$) exists between both classifications. This is important because the $[TiO]_{2b}$ index is not good outside the (~K3,~M5) range (cf. Figure 3.13 from Kenyon 1986), and we have to rely there more strongly on the visual classification. Much the same applies to the remaining indices given above: $[TiO]_1$ and $[TiO]_{2a}$ are reliable inside the spectral type interval ~K4 < ST < ~M6 (cf. Figure 2 and 3 from Kenyon and Fernandez-Castro, 1987), while [VO] is reliable in range ~M2 < ST < ~M6 (cf. Figure 4 from Kenyon and Fernandez-Castro, 1987).

Although AS 210 is classified as G: in Kenyon (1986), and the $[TiO]_{2b}$ index in Table I gives an M3 spectral type for it, it has been proven that this symbiotic system contains a carbon Mira (see, Whitelock, 1987; Schmid and Nussbaumer, 1993). This can easily be seen by noting the presence of $C_2$ Swan band $\lambda\lambda$ 5165 Å (and possibly $\lambda\lambda$ 5636 Å) and CN bands $\lambda\lambda$ 7945, 8125 and 8320 Å in Figure A.4.

Consequently, we have also classified AS 210 following the quantitative scheme of Cohen (1979), and calculated the corresponding temperature index:

$$T = \frac{2 \times I\ (5893)}{[\ I\ (5893) + I\ (6032)\ ]} \cong 0.86 - 0.87 \tag{7}$$

(note that we have slightly changed the wavelength at which we anchor the continuum in order to avoid emission lines) and the Swan bands of $C_2$ indices:

$$C_1 = \frac{I\ (5165)}{I\ (5185)} = 0.48 \tag{8a}$$

$$C_2 = \frac{I\ (5635)}{I\ (5660)} = 0.67: \tag{8b}$$

These indices imply a spectral type $C_{1,2}$ for AS 210 and $T_{eff} \approx 3500$ K ($\log T_{eff} \approx 3.54$), which is not very different from the value previously obtained from the $[TiO]_{2b}$ index. However, if we apply the qualitative classification scheme of Richer (1971), the well marked band heads at λλ 7852, 7876 and 7899 Å suggest a C3 class which, in this context, is equivalent to $T_{eff} \approx 2700$ K.

In Figure 2 we show a histogram representing the distribution of symbiotic stars among the different spectral types. The rapid increase of the number of symbiotic systems with spectral type is evident. This can be due to either the IMF relevant to the symbiotic binaries, the stellar population of the bulge, some evolutionary property favoring the strongest winds in the coldest giants, mass dependence of the residence time in the AGB, or even a selection effect. In fact, regardless of the other factors, we think that there is a selection effect at play which favors the detection of the emission spectrum in the systems having the coldest giants. To show this, we also plot in Figure 2 the ratio between Hα flux and the interpolated continuum at the same wavelength for the symbiotic stars of our sample together with a log-linear fit to the points. We can clearly see that, in fact, Hα intensity relative to the continuum increases monotonously with spectral type. In other words, it would be easier to missclassify a symbiotic as a normal giant (no line-emission spectrum) in earlier spectral types where the emission lines could be hidden in the noisy continuum of the giant star. This behavior may be due to the larger mass loss in colder (more evolved) giants and the corresponding density increase of the ionized nebula, and/or to the higher temperature of the ionization source due to the larger accretion rate from the denser wind. Calculations from our ionization model for symbiotic stars (see, Medina Tanco and Steiner, 1994) show that, under typical conditions, a variation of the hot source temperature from $5 \times 10^4$ K to $3 \times 10^5$ K would modify the Hα luminosity by only ~0.1 dex, while an order of magnitude variation in wind density can account for ~1 dex in Hα luminosity. Therefore, we think that it is the increase in density of the ionized region, due to the larger mass loss in colder giants, the responsible for the aforementioned selection effect.

## III. HR diagram

Once the spectral type of the giant stars in the binary symbiotic systems is known, and a reasonable guess can be made about their distance to the sun, an HR diagram can be constructed to get some insight into the evolutionary status of the cold component.

The first problem to address is that of transforming spectral type into effective temperature. Unfortunately, this is not a very well understood step for a cool giant star (Garrison 1994, Corbally 1994), and it is expected to be even more obscure for a giant star whose extended atmosphere is closely related to a nearby ionizing source, as is the case in a symbiotic binary system. Prior to the work of Ridgway et al. (1980), the spectral type-effective temperature calibration of Johnson (1966) was commonly used. The latter, based on the measurement of angular stellar diameters for only eight objects, is too cool for the latest type giants when compared to the calibration of Ridgway et al. (1980), based on measurements of 31 giant stars.

In Figure 3 we reproduce figure 3 of Ridgway et al. (1980), where their calibration and that of Johnson (1966) are compared. We also include in this figure the original data used by Ridgway et al. (1980) in deriving their calibration (from tables 2 and 3 of that work). The numbers beside the dots correspond to the number of independent determinations of each data point. It can be seen that the dispersion of the points is so large that it can be even asked whether a one-to-one relation between spectral type and effective temperature actually exists for cold giants. In particular, it can be seen that a calibration as cold as that of Johnson (1966) cannot be disregarded. At any rate, the calibration relation for a symbiotic giant should be equal to, or cooler than, the corresponding relationship for an isolated giant. This is so because of the presence of the hot companion which could shift systematically the inferred spectral type of the giant companion to earlier types.

Considering these uncertainties we discuss our results for both, Johnson (1966) and Ridgway et al. (1980) calibrations.

Given $T_{eff}$, the K magnitude and the distance to the objects, some knowledge is needed about the spectral distribution of the energy emitted by them in order to calculate their bolometric luminosities. The presence of strong molecular absorption bands makes this a difficult step for late type objects. The safest way to obtain reliable luminosities is to use bolometric corrections, $BC_K$, for the K magnitude of bulge giants. An alternative, but more dangerous way, is to assume a black body emission spectrum at $T_{eff}$. Unfortunately, $BC_K$ are color dependent and there are color determinations for only a few objects of our sample. Therefore, we will apply both methods: first, the black body approximation for all the objects and, second, $BC_K$ corrections for those objects for which $(J-K)_0$ is available.

If we assume that the spectrum of the giant component can be well represented by a black body at $T_{eff}$, then the luminosities can be evaluated from the K magnitudes of the objects. This requires, of course, the knowledge of the distance, which we take as the distance to the galactic center, *d=8.5* kpc, since all the sources in the sample are located in the direction of the galactic center ( $(l^2+b^2)^{1/2} \leq 20°$ ). The K magnitudes used here are those listed in Kenyon (1986), corrected due to galactic extinction according to the expression (c.f., Whitelock, 1987; Jaschek and Jaschek, 1987):

$$A_K = 0.123 \times 0.099 \times \left( \frac{1}{\sin|b|} - 1 \right) \times \left[ 1 - \exp(-10 \cdot d \cdot \sin|b|) \right] \qquad (9)$$

where *b* is the galactic latitude and *d* is the distance to the object. We do not introduce further corrections for circumstellar reddening. An approximate treatment for circumstellar extinction could be that of Whitelock (1987) for D-type symbiotics, which seems to point to a typical circumstellar extinction *$A_K$~0.6* in these dusty objects. This effect is certainly of importance in D-type symbiotics, which are very likely Mira stars enshrouded by a thick shell of dust at some few hundred degrees Kelvin. Nevertheless, for S-type symbiotics, which constitute *~90%* of our sample, we do not expect circumstellar extinction to play an important role. Therefore, we neglect this correction and limit ourselves only to the inclusion of interstellar extinction by means of equation (9), noting that the luminosities of the five D-type symbiotic stars of our sample are probably underestimated by one fourth of a decade.

Luminosities ($L/L_\odot$), effective temperatures ($T_{eff}$ - from Ridgway's calibration) and absolute K magnitudes ($M_K$), calculated as explained above, are given in Table III. H-R diagrams for the symbiotic stars under study are shown in Figures 4.a and 4.b for the spectral type-$T_{eff}$ calibrations of Johnson (1966) and Ridgway et al. (1980) respectively. In these figures, empty circles correspond to S-type symbiotics, and filled circles correspond to D-type objects. Actually, each point corresponds only to the cool giant component of the system, since the other components, mainly the hot ionizing source and circumstellar dust, make a negligible contribution at K wavelengths, and the same applies to the 6000 Å - 8000 Å interval where spectral type classification was done. The error bars shown in the figures are only indicative of the uncertainty introduced by the spectral type classification. A more important uncertainty is that introduced by the translation of spectral type into effective temperature for a cool giant, which can be illustrated through the comparison of both figures 4a-b.

Evolutionary trajectories of intermediate mass stars are also displayed in Figures 4a-b. The tracks for *M=3, 4, 5, 7* and *9 $M_\odot$* are from Castellani, Chieffi and Straniero (1990), and those for *M=1, 1.5* and *2.25 $M_\odot$* were taken from Seal (1990) and are due to Iben and Tutukov (1984). All the trajectories correspond to metal rich single stars (*Y=0.27* and *Z=0.02*) and were computed from the main sequence up to the full H-re-ignition, which identifies the onset of the thermally pulsating phase. However, we linearly fit straight lines to the last portion of each AGB of tracks with $M \geq 1.5\ M_\odot$ as an approximation to the thermally pulsating phase of the AGB (**TP-AGB**). On these tracks we superimpose the location of the first thermal pulse as given by the synthetic model of Groenewegen and Jong (1993), and the end of the AGB (the onset of the superwind) calculated from Iben and Renzini (1983).

It can be seen that most giant components of symbiotic systems can be adjusted to $M \in (\sim 1.5\ M_\odot, \sim 4\ M_\odot)$ tracks, although masses as high as *7 $M_\odot$* cannot be discarded. Nevertheless, the giants seem to have preferentially $M < 4\ M_\odot$ clustering around $M \approx 2\text{-}3\ M_\odot$ depending on the ST-$T_{eff}$ relation used. This is somewhat lower than the value proposed by Seal (1990) while, at the same time, no clear match with a single evolutionary track can be seen. This is more noticeable for the cool ST-$T_{eff}$ relation of Johnson.

There are a few objects with positions beyond the theoretical end of the AGB, but they are probably there due to overestimation of the distance *d* (e.g., AS 270, AS 289 and V2416 Sgr) or wrong spectral type classification (e.g., H2-5 and V455 Sco).

More troublesome are the four hottest objects of the sample (from bottom to top parallel to the AGBs in 4a: Hen 1591, Hen 1410, AS 255 and SS 129). An error in temperature classification could push them down and to the right from the *7-9 $M_\odot$* track into the $\sim 1.5\ M_\odot, \sim 3\ M_\odot$ interval, in the non thermally-pulsating (E-AGB) portion of the AGBs. However, excepting Hen 1410, there is excellent agreement in spectral type classification of these objects between Kenyon and us, and so they seem to be actually earlier type giants[3]. Hence, the solution to the problem does not seem to be modifying $T_{eff}$. Another possibility is that their distances were overestimated. In this case, they could be moved down at constant temperature up to the AGBs of *3-5 $M_\odot$* stars, or the RGB of stars of $M_* < 3\ M_\odot$. Now, in order to avoid confusion, we name as A-type objects those giant companions located along the E-AGB, and B-type objects those along the TP-AGB. Then, if the four hot objects had their distances overestimated and belong actually to the foreground disk, they are A-type objects. Assuming homogeneity, a simple geometrical argument implies that, if there are *4* such objects in the foreground, there must be ~*12* of them inside the bulge. As the distance to this objects would have been correctly assumed, they should appear along the E-AGB portion of the evolutionary tracks in our HR diagram. As they are not, there are two possibilities: either there are not A-type objects in the bulge or, if there are, we are not able to detect them as symbiotic stars. The latter could probably be achieved by the selection effect described in the previous section. If this were true, and A-type objects are progenitors of B-type stars, then we should observe B-type objects inside the same solid angle but in the foreground. These high luminosity objects would have had their distances overestimated and, consequently, they should appear in our HR diagram as high mass high luminosity objects. Finally, the most likely explanation for the position of these warm objects in the HR diagram could be low metallicity (see Figure 6 and text below).

---

[3] Though it cannot be ensured that the $T_{eff}$ obtained from the spectral type is completely meaningful for giants as late as the cool components of symbiotic binaries.

The cool components of D-type symbiotics are thought to be extreme Mira stars (e.g., Whitelock 1987), or short period OH/IR stars (e.g., Schild 1989), with high mass loss rates and near the onset of a superwind regime which will eventually lead to the ejection of a planetary nebula. This is in fact the picture arising from Figures 4a and 4b, regardless of the ST-$T_{eff}$ transformation relation used. In both cases the symbiotic phenomenon seems to develop only inside the thermally pulsating strip, being triggered immediately after (or near - for sufficiently high masses) the first thermal pulse. The D-type stars (filled circles) are the most evolved objects along each evolutionary track when compared to S-type objects of similar mass. Furthermore, they appear only after the beginning of the thermally pulsating phase and tend to be located in the vicinity of the end of the AGB. If the actual ST-$T_{eff}$ is as hot as that of Ridgway et al., then there seem to be no *visible* symbiotic stars near the onset of the superwind, probably because by this time the binary is obscured by the thick ejecta of the giant.

The question arises as to whether S and D-type symbiotics are different objects or different evolutionary stages of the same object. In other words, were D symbiotics S objects at some time in the past ? We can use the *Fuel Consumption Theorem* (Renzini and Buzzoni, 1986) to try to answer this question. It can be written as:

$$N_j = B_j(t) \cdot L_T \cdot t_j \tag{10}$$

where $N_j$ is the number of objects in the evolutionary stage $j$, of duration $t_j$, $L_T$ is the total luminosity of the population and $B_j(t)$ is the evolutionary flux, which can be taken as constant ($B_j(t) \approx 2 \times 10^{-11}$ * / yr / $L_\odot$). If we assume that every S-type object in the bulge goes through a D-type phase, taking $L_T = 4.9 \times 10^5\ L_\odot$ as the total luminosity of the symbiotic stars in the bulge (from Table III), and the lifetime of a D-type system as that of the Mira phase, i.e., $t_j \approx 4 \times 10^4$ yr, we get an expected number of D-type symbiotics in the bulge $N_D \approx 0.4$. This number is a factor of *10* smaller than the number of D symbiotics we actually know in the bulge. Therefore, either we know at most *10 %* of the S-type symbiotics in the bulge (the remaining missidentified as planetary nebula ?) but all the D-type systems, or somehow a Mira manages to survive ten times longer when forming part of a symbiotic binary than when it is isolated (difficult to accept), or, finally, S-type and D-type symbiotics keep no genetical relationship. We think the latter option is the most likely.

We note also that Seal (1990), in his HR diagram obtains a clear separation between S and D-type objects. There, the D-type symbiotics are the latest type objects, with *log $T_{eff}$* ≤ 3.45. We do not see the same effect. Only one of our D symbiotics, H2-38 (and, possibly, AS 210 - c.f., Section II), falls in such a cold region of the HR diagram. The remaining D-type symbiotics share almost the same effective temperature region than the S-type objects, the only systematic difference arising in luminosity (and, if they are actually later type objects, they are so only in the sense of pointing toward a more evolved stage).

The previous results were obtained under the (strong) assumption of black body emission for the cold giant stars. However, the best way to calculate the actual luminosities of the giant stars from K measurements is to use the corresponding bolometric corrections, $BC_K$. Frogel and Whitford (1987), give $BC_K$ as function of $(J-K)_0$ for M giants in the Baade's window (see their Figure 1). Unfortunately, they only cover the interval *0.6 < $(J-K)_0$ < 1.5*, while the few objects for which we have observed values of (J-K) (from Munari et al., 1992) extend over the interval *1.0 < $(J-K)_0$ < 3.1*, and are mainly centered in *1.0 < $(J-K)_0$ < 2.0*. To circumvent this problem, we calculated the bolometric corrections $BC_K$ as a function of $(J-K)_0$ for Mira stars in the galactic bulge from the data given in Table 3 of Whitelock, Feast and Catchpole (1991), and used them to extend the original figure of Frogel and Whitford (1987) up to $(J-K)_0 \cong 5$. The results of a 5th order polynomial fit to the whole sample of M giants plus Mira stars is given in Figure 5. We note that although the right fit is probably given by a curve going through the M giants, but above the Mira stars, as the latter have their K bands depleted by circumstellar dust absorption, the difference is of no significance in our case.

The luminosities calculated in this way are plotted in the HR diagram of Figure 6, where now the evolutionary tracks are from Schaller et al. (1992) (for *M = 1 and 2 $M_\odot$* and Z=0.001 and *M = 1, 1.25, 1.5, 2 and 3 $M_\odot$*, and Z=0.02), and Schaerer et al. (1993) (for *M = 1, 1.25, 1.5, 2 and 3 $M_\odot$*, and Z=0.04)[4]. We can see that our results do not change because of the different luminosity determination, even if the dispersion of the points is somewhat increased. The evolutionary tracks were again calculated up to the end of the early-AGB, beginning of the TP-AGB, and the objects are still inside the TP-AGB region, regardless of the ST-$T_{eff}$ relationship used. Furthermore, we can see that the giant companions can span a considerable metallicity range, though more detailed conclusions depend heavily on the ST-$T_{eff}$ relationship assumed. For a Johnson's (cool) transformation relation we can see that most of the objects must be of high metallicity (Z=0.02-0.04 and larger), though some few very low metallicity objects could exist, which could help to explain the positions of Hen 1591, Hen 1410, AS 255 and SS 129 in the HR diagram of Figure 4a. The right-most track in Figure 6, corresponds to *M = 1 $M_\odot$* and *Z = 0.04*, and an object at its tip should be *$1.2 \times 10^{10}$* yr old. The corresponding age for a, say, *M = 0.8 $M_\odot$*, *Z = 0.04* object would be *> $2.7 \times 10^{10}$* yr, and so the objects at the right of the extrapolation of the *M = 1 $M_\odot$* track into the TP-AGB region cannot be explained by much lower mass progenitors, and so those giants might have metalicities *Z > 0.04*. For Ridgway's et al. transformation, on the other hand, there is no need for metallicities in excess of solar values, though very low metallicity objects could still be present in the sample.

We note that, although not shown, the luminosities have also been calculated using the K-dependent $BC_K$ used by Rich, Mould and Graham (1993) for the case of M31 giants, and even a constant bolometric correction, $BC_K = 3.1$, as used by Glass (1993) for non-Mira stars of the galactic bulge, and the results are qualitatively unchanged.

Finally, in Figure 7 we show a magnitude versus color diagram, $K_0$ vs. $(J-K)_0$, for both M giants and Mira stars in the galactic bulge (these data are from Frogel and Whittford, 1987 and Whitelock, Feast and Catchpole, 1991, respectively), and those symbiotic stars of our sample for which $(J-K)_0$ is available. It can be seen that the symbiotic giants follow the same trend as the other single giants, and are strongly concentrated near and at the tip of the AGB. There are only three uncommonly luminous objects, AS 289, AS 276 and V2416 Sgr, which could probably be foreground objects belonging to the galactic disk whose distances were overestimated.

---

[4] The evolutionary tracks from Schaller et al. (1992) and Schaerer et al. (1993) were used this time because they are newer than those used in Figure 3 and include many improvements not consider previously. New radiative opacities by Rogers and Iglesias (1991) have been included. These differences with respect to the Los Alamos 1977 opacities typically amount to a factor 3 at $3 \times 10^5$ K for solar metallicity. Many outputs of the evolutionary models are modified, like surface parameters (L, R, $T_{eff}$), envelope structure and pulsation properties. In any case, the differences between the models are not relevant for the present work.

## IV. Conclusions

We made spectroscopic observations of 45 symbiotic stars in the direction of the bulge of the galaxy. The spectra were classified in temperature (spectral type) according to TiO $\lambda$ 6180, and 7100 Å and VO $\lambda$ 7865 Å absorption feature indices as well as visual comparison with standard spectra. The results of this classification, together with K data available in the literature, were used to estimate the luminosity of the giant stars both, under the assumption of black body continuum emission at $T_{eff}$ and through $BC_K(J-K)$ for bulge giants. The final results were used in the construction of an HR diagram where the symbiotic stars are superposed on evolutionary tracks of intermediate mass stars. A magnitude versus color diagram, $K_0$ vs. $(J-K)_0$, gathering bulge M giants, Miras and symbiotic stars is also shown. The main results arising from this work are summarized bellow:

1) The number of symbiotic stars increases towards later spectral types. However, this could be the result of a selection effect favoring the identification of the symbiotic phenomenon over later type giant spectra, when the emission spectrum is relatively more intense.

2) The mass of the giant components seem to be scattered in the interval (*~1.5 - ~4 $M_\odot$*), with a certain preference towards the low mass end. Nevertheless, masses as high as *~7 $M_\odot$* cannot be discarded.

3) D-type symbiotics appear as more evolved objects than S-type symbiotics of similar mass, and are located near the end of the AGB. It is not easy to state whether they are different objects, or different evolutionary stages of the same object. However, the application of the *Fuel Consumption Theorem* points towards no genetical relationship between S and D symbiotics.

4) Almost all the giant components in symbiotic stars, regardless of whether they are S or D-type, lie inside the thermally pulsating phase of the evolutionary tracks. So, even if the wind is present along the whole AGB, before the first thermal pulse it is probably not strong enough to feed the hot accreting source to heat it up to the high temperatures required for the symbiotic phenomenon to develop. If this is the case, then the appearance of the symbiotic stars only after the beginning of the thermally pulsating phase is a strong argument for a close relationship between thermal pulses and mass loss in AGB. There must be a systematic difference in wind intensity between the E-AGB and the TP-AGB, and the variation between both regimes must be rather abrupt. This seems to favor a scenario of the kind proposed by Baud and Habing (1983) and Bedijn (1988) (see also Habing, 1989), where an *accelerated* mass loss exists which increases slowly at first until an exponential regime is reached.

5) The giant components of symbiotic stars in the bulge seem to span a considerable range of metalicity, covering from very low Z up to perhaps more than twice solar for a cool enough spectral type-effective temperature transformation relation.

Finally, even if most of the previous results are qualitatively independent of the ST-$T_{eff}$ relation used, we stress that the importance of a reliable ST-$T_{eff}$ relation for the understanding of the evolutionary status of symbiotic giants cannot be overemphasized.

## Acknowledgments

The authors benefited from useful suggestions from R. Michael Rich. G. A. Medina Tanco thanks the Brazilian agency **FAPESP**, whose financial support under grant 92/4517-0 made this work possible.

# References

Allen D. A. (1984). MNRAS, **192**, 591.

Allen D. A. (1984). Proc. Astron. Soc. Aust., **5**, 369.

Altamore A., Baratta G. B., Casatella A., Friedjung M., Giangrande A. and Viotti R., (1981). Ap. J., **245**, 630.

Baud B. and Habing H. J., 1983: Astron. Astrophys., **127**, 73.

Bedijn P. J., 1988. Astron. Astrophys., **186**, 136.

Castellani V., Chieffi A. and Straniero O., (1990). Astrophys. J. Suppl. Ser., **74**, 463.

Cohen M., 1979. MNRAS, **186**, 837.

Corbally C. J., (1994). Personal communication.

Frogel J. A. and Whitford A. E., (1987). Ap. J., **320**, 199.

Garrison R., (1994). Personal communication.

Glass I. S., (1993). in *Galactic Bulges*, H. Dejonghe and H. J. Habing (eds.), Kluwer Academic Press (Printed in the Neatherlands), pp. 21-37.

Groenewegen I. A. and de Jong T., (1993). Astron. Astrophys., **267**, 410.

Habing H. J., 1989. In *From Miras to Planetary Nebulae: Wich path for Stellar Evolution*, M. O. Mennessier and A. Omont (eds.), Editions Frontieres, pp 16-40.

Iben I. Jr. and Tutukov A. V., (1984). Astrophys. J. Suppl. Ser., **54**, 335.

Iben I. Jr. and Renzini A., (1983). Ann. Rev. Astron. Astrophys., **21**, 271.

Iglesias C. A.and Rogers F. J., 1991. Ap.J., **371**, 408.

Jacoby G. H., Hunter D. A. and Christian C. A., (1984). Ap. J. Suppl. Series, **56**, 257.

Jaschek C. and Jaschek M., (1987). *The Classification of Stars,* Cambridge University Press, Cambridge.

Johnson H. L., (1966). Ann Rev. Astron. Astrophys., **4**, 193.

Kenyon S. J., (1986). The Symbiotic Stars (Cambridge University, Cambridge).

Kenyon S. J., (1988). Astron. J., **96**, 337.

Kenyon S. J. and Webbink R. F., (1984). Ap. J. ,**279**, 252.

Kenyon S. J. and Fernandez-Castro T., (1987). Astron. J., **93**, 938.

Kenyon S. J., Fernandez-Castro T. and Stencel R. E., (1988). Astron. J., **95**, 1817.

Medina Tanco G. A. and Steiner J. E., (1994). Submitted for publication.

Michalitsianos A. G., Kafatos M. and Hobbs R. W., (1980). Ap. J., **237**, 506.

Munari U., Yudin B. F., Taranova O. G., Massone G., Marang F., Roberts G., Winkler H. and Whitelock P. A., (1992). Astron. Astrophys. Suppl. Ser., **93**, 383.

Renzini A. and Buzzoni A., 1986. in *Spectral Evolution of Galaxies*, C. Chiosi anmd A. Renzini (eds.), Dodrecht: Reidel, p. 153.

Rich R. M., Mould J. R. and J. R. Graham, (1993). Astron. J., **106**, 2252.

Richer H. B., 1971. Ap. J., **167**, 521.

Ridgway S. T., Joyce R. R. and Wing R. F., (1980). Ap. J., **235**, 126.

Seal P., (1990). Astrophys. Sp. Sci., **174**, 321.

Schaerer D., Charbonnel C., Meynet G., Maeder A., Schaller G., (1993). Astron. Astrophys. Suppl. Ser., **102**, 339.

Schaller G., Schaerer D., Meynet G., Maeder A., (1992). Astron. Astrophys. Suppl. Ser., **96**, 269.

Schild H., (1989). MNRAS, **240**, 63.

Schmid and Nussbaumer, 1993. Astron. Astrophys, **268**, 159.

Schulte-Ladbeck R. E., (1988). Astron. Astrophys., **189**, 97.

Stencel R. E., Michalitsianos A. G., Kafatos M. and Boyarchuk A. A., (1982). Ap. J. (Letters), **253**, L77

O'Connell R. W., (1973). Astron. J., **78**, 1074.

Whitelock P. A., (1987). Pub. Astron. Soc. Pacific, **99**, 573.

Whitelock P. A., Feast M., and Catchpole R., (1991). MNRAS, **248**, 276.

**Figure Captions:**

**Figure 1:** relationship between $[TiO]_{2b}$ classification and the final visual classification (columns 4 and 5 of Table I).

**Figure 2:** the number of stars (left axis), S+D-type together, and ratio between $F(H\alpha)/F_c(H\alpha)$ (right axis), where $F_c$ is the flux at the interpolated continuum, as a function of spectral type. The latter quantity is expected to represent the intensity of the emission spectrum relative to the continuum intensity in symbiotic stars. The straight line is a log-linear fit to the points.

**Figure 3:** spectral type-effective temperature transformation relations for giant stars by Ridgway et al. (1980) (heavy line) and Johnson (1966) (light line). The dots with their error bars are the original data used by Ridgway et al. (1980) in deriving their relation, while the numbers by the dots are the number of independent determinations of each data point. (This figure is an adaptation of figure 3 and tables 2 and 3 of Ridgway et al (1980).)

**Figure 4a-b:** HR diagram for symbiotic stars. empty circles correspond to S-type objects, and filled symbols to D-type objects. $T_{eff}$ is obtained from our own spectral type classification, and $L/L_\odot$ is obtained from K values given by Kenyon (1986), corrected for interstellar extinction using a galactic latitude dependent formula, and under the assumption of black body emission at d=8.5 kpc. The evolutionary tracks for M=3,4,5,7 and 9 $M_\odot$ are from Castellani et al. (1990), while the remaining tracks are from Iben and Tutukov (1984). The evolutionary tracks were originally calculated up to the onset of the thermal pulses, and we have extrapolated them to later stages by fitting a straight line to the last portions of the AGBs. The location of the first thermal pulse is shown as calculated from the synthetic models of Groenewegen and Jong (1993), while the end of the AGB also displayed is from Iben and Renzini (1983). The evolutionary tracks are for metal rich stars (Y=0.27 and Z=0.02). In **(a)** $T_{eff}$ from Johnson (1966), while in **(b)** $T_{eff}$ from Ridgway et al. (1980). The error bars take into account only the propagation of the spectral type classification error, while the uncertainty in ST-$T_{eff}$ transformation can be evaluated by comparing figures (a) and (b).

**Figure 5:** Bolometric correction, $BC_K$, for K magnitude. The crosses correspond to M giants in the Baade's window and were taken from Figure 1 of Frogel and Whitford (1987). The open circles correspond to Mira stars in the galactic bulge and were calculated from Table 3 of Whitelock, Feast and Catchpole (1991). The solid curve is a 5th order polynomial fit to the points.

**Figure 6:** Symbiotic stars for which $(J-K)_0$ is known (from Munari et al., 1992) and whose luminosities were calculated from the $BC_K$ obtained from Figure 5. Evolutionary tracks for *M = 1, 1.25, 1.5, 2 and 3 $M_\odot$* with Z=0.02 (light lines), and 0.04 (heavy lines) and *M = 1 and 2 $M_\odot$* with Z=0.001 (heavy lines indicated in the figure) from Scheller et al. (1992) and Schaerer et al. (1993). All the tracks end at the beginning of the TP-AGB. Two sets of points are given: one for Johnson(1966) (light stars) and another for Ridgway's et al. (1980) (heavy stars) ST-$T_{eff}$ transformation. Error bars are not given to avoid the pollution of the figure, but the errors in temperature due to ST classification are the same as in figures 4a-b, while the errors in luminosity should be considerably smaller. The uncertainty in ST-$T_{eff}$ transformation can be estimated by comparing both sets of points.

**Figure 7:** Magnitude vs. color diagram, $K_0$ vs. $(J-K)_0$ for bulge M giants (Frogel and Whitford, 1987), bulge Miras (Whitelock et al., 1991) and bulge symbiotic stars of our sample.

**Figures A.1-A.45:** observed spectra of the symbiotics in the direction of the galactic bulge and used in this work.

**Table Caption for Table I**

**Table I:** Objects observed and spectrally classified in this work. Column 3 is the classification given in Kenyon (1986); column 4 is the result of using [TiO]$_{2b}$ (two values separated by a slash correspond to spectra obtained in different nights) and the last column is the final (visual) classification.

# Table I.

| Object | IR-Type | Kenyon (1986,1988) | $[TiO]_{2h}$ λ7100 | M.T.&S. |
|---|---|---|---|---|
| **Wra 1470** | S | M4 | | M6 |
| **He2-171** | D | M | M4.7 | M6 |
| **He2-173** | S | M | M6.2 | M7 |
| **AS 210** | D | G: | M3.0 | M3: / $C_{1,2}$ |
| **HK Sco** | S | M | | M1 |
| **CL Sco** | S | M | | M2 |
| **V455 Sco** | S | M6 | M6.9 | M6 |
| **Hen 1342** | S | M2 | K5.4/5.6 | K5-M0 |
| **AS 221** | S | M4 | | M4 |
| **H2-5** | S | M | M5.7 | M7 |
| **Th3-7** | S | M | M5.6 | M5-6 |
| **Th3-18** | S | M2 | | M3 |
| **Hen 1410** | S | M | | K2 |
| **Th3-30** | S | K5 | | K5 |
| **AE Ara** | S | M2 | M5.7 | M5 |
| **SS 96** | S | M2 | M2.7 | M0 |
| **H1-36** | D | M | M4.3 | M4-5: |
| **AS 255** | S | K3 | K3.0 | K3 |
| **V2416 Sgr** | S | M5 | | M5 |
| **SS 117** | S | M6 | | M5 |
| **Ap1-8** | S | M0 | | M2 |
| **SS 122** | S: | M7 | | M7 |
| **AS 270** | S | M1 | | M2 |
| **H2-38** | D | M8 | | M7-8 |
| **SS 129** | S | K | K4.2 | K3 |
| **V615 Sgr** | S | M | | M5 |
| **Hen 1591** | S | G | K3.1 | K1 |
| **AS 276** | S | M4 | M3.5 | M3-4 |
| **Ap1-9** | S | K4 | K5.2 | K5 |
| **AS 281** | S | M5 | M4.3 | M5 |
| **V2506 Sgr** | S | M | M3.8 | M4 |
| **SS 141** | S | M | M4.5 | M4 |
| **AS 289** | S | M3 | M3.8 | M3-4 |
| **Y CrA** | S | M | M6.0 | M6 |
| **V2756 Sgr** | S | M | M2.6 | M3 |
| **HD 319167** | S | M3 | | M3 |
| **He2-374** | S | M | M5.1 | M5 |
| **Hen 1674** | S | M5 | M4.2 | M5 |
| **He2-390** | D | M | M1.4 | M1-2 |

| Object | IR-Type | Kenyon (1986,1988) | $[TiO]_{2h} \lambda 7100$ | M.T.&S. |
|---|---|---|---|---|
| **V3804 Sgr** | S | M6 | | M5 |
| **V3811 Sgr** | S | M | K5.5 | K5-M0 |
| **V2601 Sgr** | S | M | | M2 |
| **AS 316** | S | M | | M3 |
| **AS 327** | S | M | M4.3 | M4 |
| **FN Sgr** | S | M | M4.4/4.1 | M4 |

**Table Caption for Table II**

**Table II:** Averaged spectral type of the sample and averaged difference in the spectral type of a given object in our sample when classified using the different absorption indices defined in (2.a-c) and (4).

**Table II**

| Feature | ⟨ST⟩ | $\sigma_{2a.i}$ | $\sigma_{2b.i}$ | $\sigma_{1.i}$ | $\sigma_{VO.i}$ |
|---|---|---|---|---|---|
| $[TiO]_{2a}$ | M3.8 | 0 | | | |
| $[TiO]_{2b}$ | M4.0 | 3.0 | 0 | | |
| $[TiO]_1$ | M4.1 | 2.8 | 1.6 | 0 | |
| $[VO]$ | M4.0 | 1.5 | 2.6 | 2.6 | 0 |

**Table Caption for Table III**

**Table III:** Effective temperature calculated from the spectral type by using Ridgway et al. (1980) for giant stars, and the corresponding bolometric luminosity calculated under the assumption of blackbody emission and *d=8.5* kpc. The last column gives the absolute K magnitude.

# Table III

| Name | $T_{eff}$ | $L/L_\odot$ | $M_K$ |
|---|---|---|---|
| **Wra_1470** | 3235 | 3693 | -6.88 |
| **He2-171** | 3235 | 16665 | -8.52 |
| **He2-173** | 3073 | 9298 | -7.97 |
| **As_210** | 3649 | 12743 | -8 |
| **HK_Sco** | 3790 | 4633 | -6.82 |
| **CL_Sco** | 3720 | 4473 | -6.82 |
| **V455_Sco** | 3235 | 24105 | -8.92 |
| **Hen_1342** | 3921 | 2808 | -6.21 |
| **AS_221** | 3560 | 5923 | -7.21 |
| **H2-5** | 3073 | 31782 | -9.31 |
| **Th3-7** | 3343 | 3306 | -6.7 |
| **Th3-18** | 3649 | 3671 | -6.64 |
| **Hen_1410** | 4428 | 4282 | -6.4 |
| **Th3-30** | 3976 | 4140 | -6.6 |
| **AE_Ara** | 3430 | 16792 | -8.42 |
| **SS_96** | 3873 | 21562 | -8.44 |
| **H1-36** | 3502 | 11623 | -7.98 |
| **AS_255** | 4256 | 3835 | -6.37 |
| **V2416_Sgr** | 3430 | 111562 | -10.47 |
| **SS_117** | 3430 | 4934 | -7.09 |
| **Ap1-8** | 3720 | 5046 | -6.95 |
| **SS_122** | 3073 | 12420 | -8.29 |
| **AS_270** | 3720 | 77875 | -9.92 |
| **H2-38** | 2992 | 10322 | -8.13 |
| **SS_129** | 4256 | 5700 | -6.8 |
| **V615_Sgr** | 3430 | 5121 | -7.13 |
| **Hen_1591** | 4609 | 2941 | -5.9 |
| **AS_276** | 3608 | 3462 | -6.6 |
| **Ap1-9** | 3976 | 2380 | -6 |
| **AS_281** | 3430 | 9492 | -7.8 |
| **V2506_Sgr** | 3560 | 2766 | -6.39 |
| **SS_141** | 3560 | 1517 | -5.73 |
| **AS_289** | 3608 | 69348 | -9.86 |
| **Y_CrA** | 3235 | 11289 | -8.09 |
| **V2756_Sgr** | 3649 | 4878 | -6.95 |
| **HD319167** | 3649 | 6376 | -7.24 |

**Table III - cont...**

| Name | $T_{eff}$ | $L/L_\odot$ | $M_K$ |
|---|---|---|---|
| **He2-374** | 3430 | 17169 | -8.44 |
| **Hen_1674** | 3430 | 4833 | -7.06 |
| **He2-390** | 3754 | 10789 | -7.76 |
| **V3804_Sgr** | 3430 | 6722 | -7.42 |
| **V3811_Sgr** | 3921 | 3101 | -6.31 |
| **V2601_Sgr** | 3720 | 4118 | -6.73 |
| **AS_316** | 3649 | 4760 | -6.93 |
| **AS_327** | 3560 | 2324 | -6.2 |
| **FN_Sgr** | 3560 | 4095 | -6.81 |

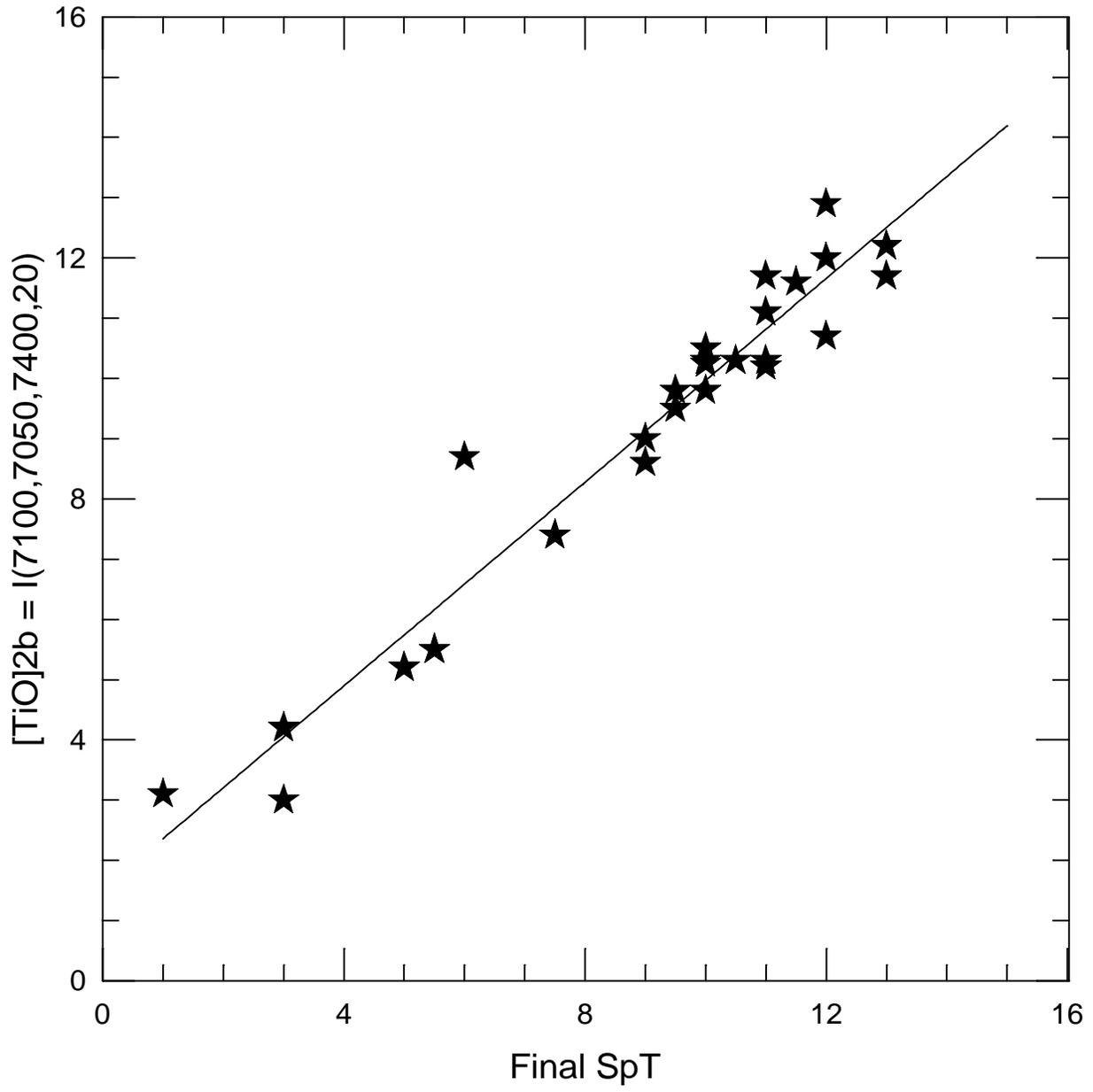

**Figure 1**

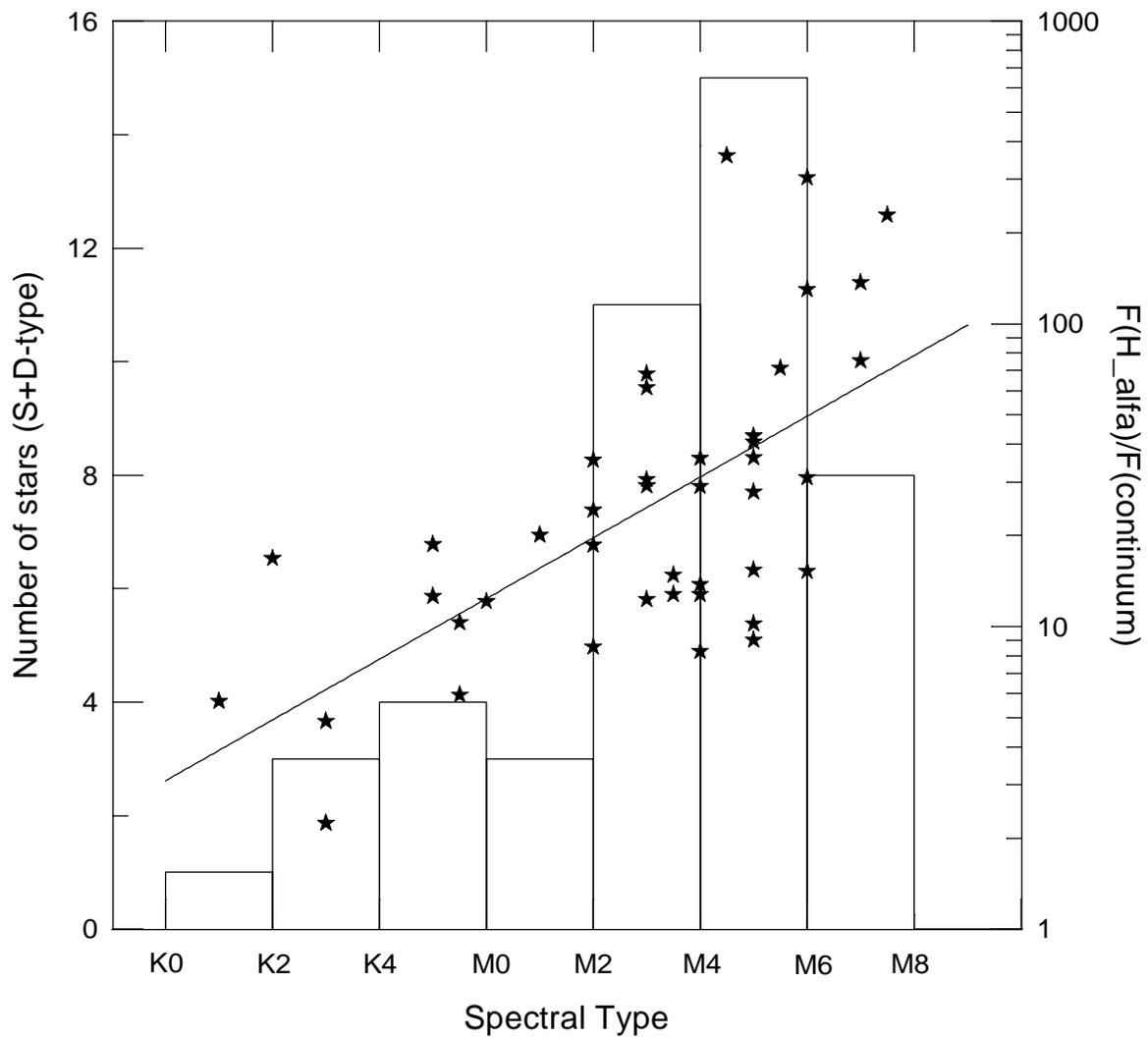

**Figure 2**

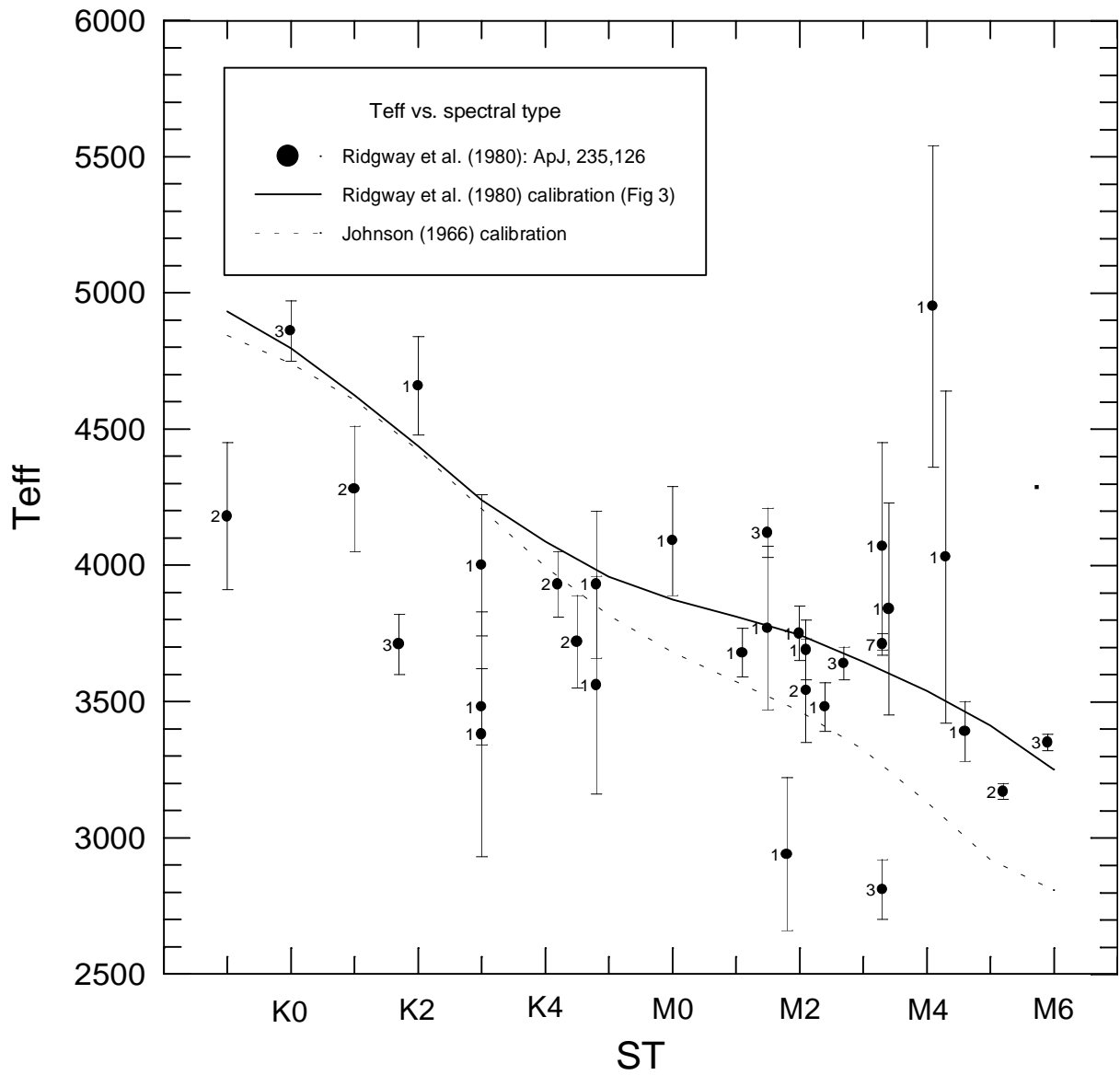

**Figure 3**

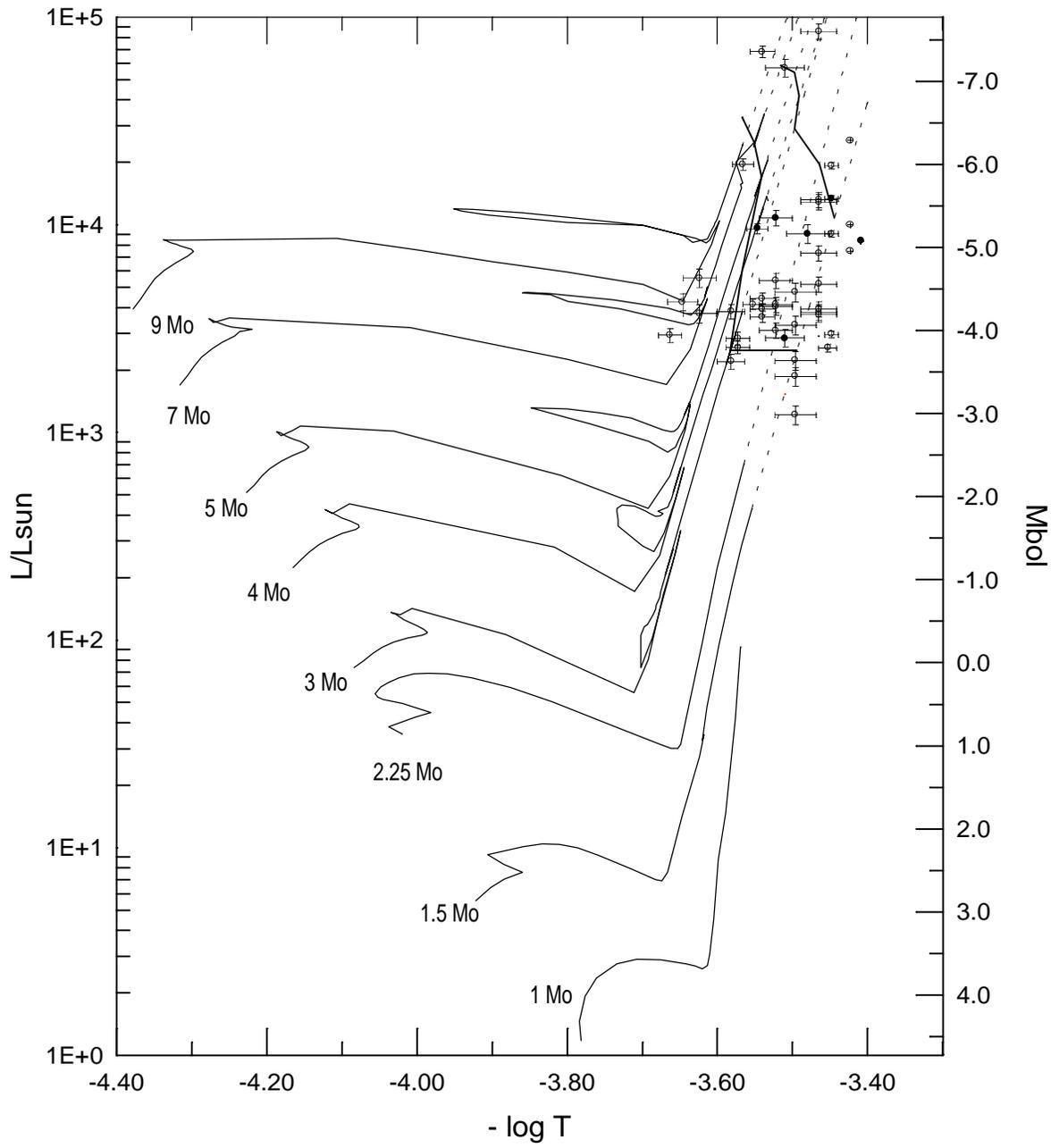

**Figure 4a**

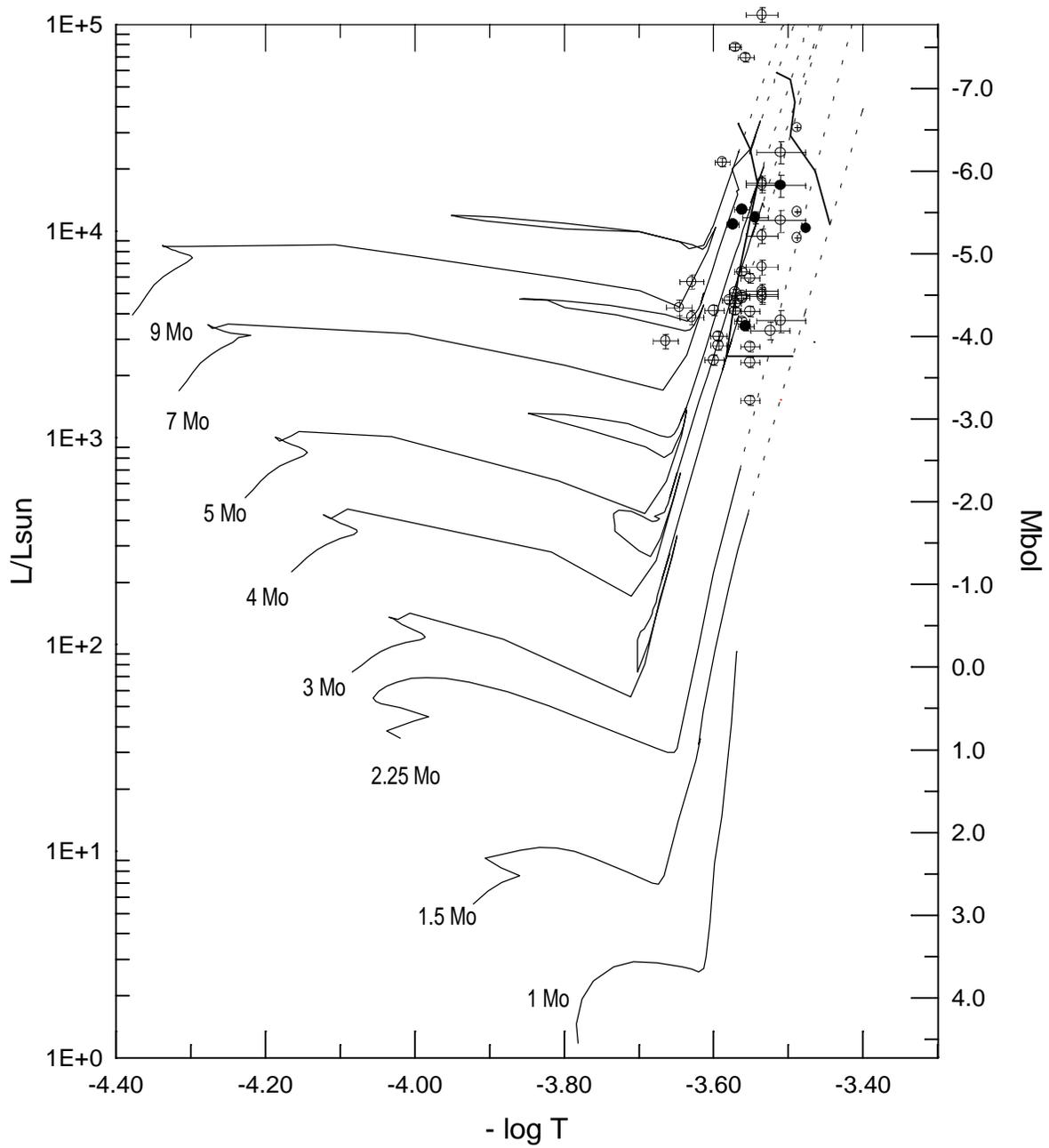

**Figure 4b**

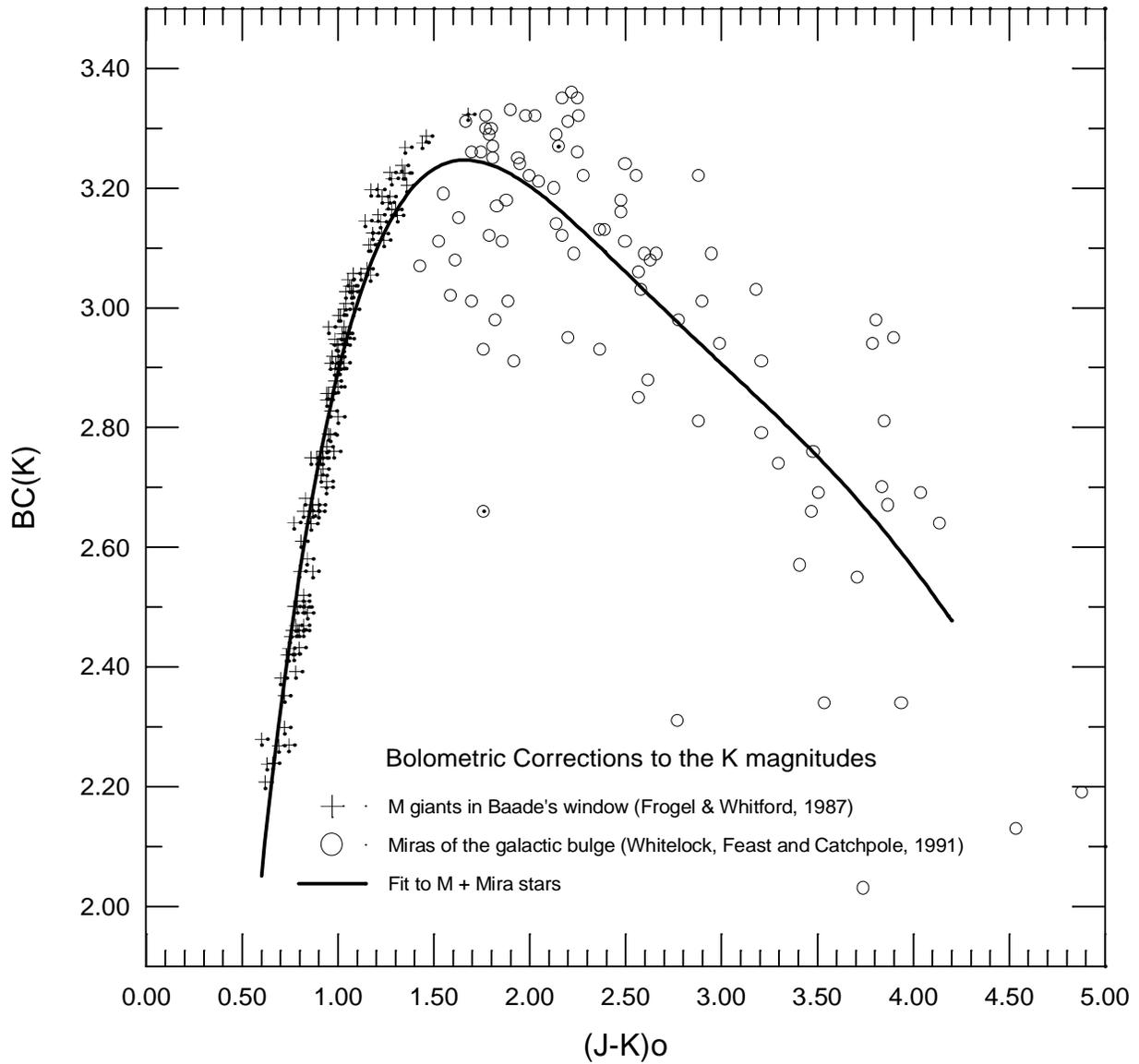

**Figure 5**

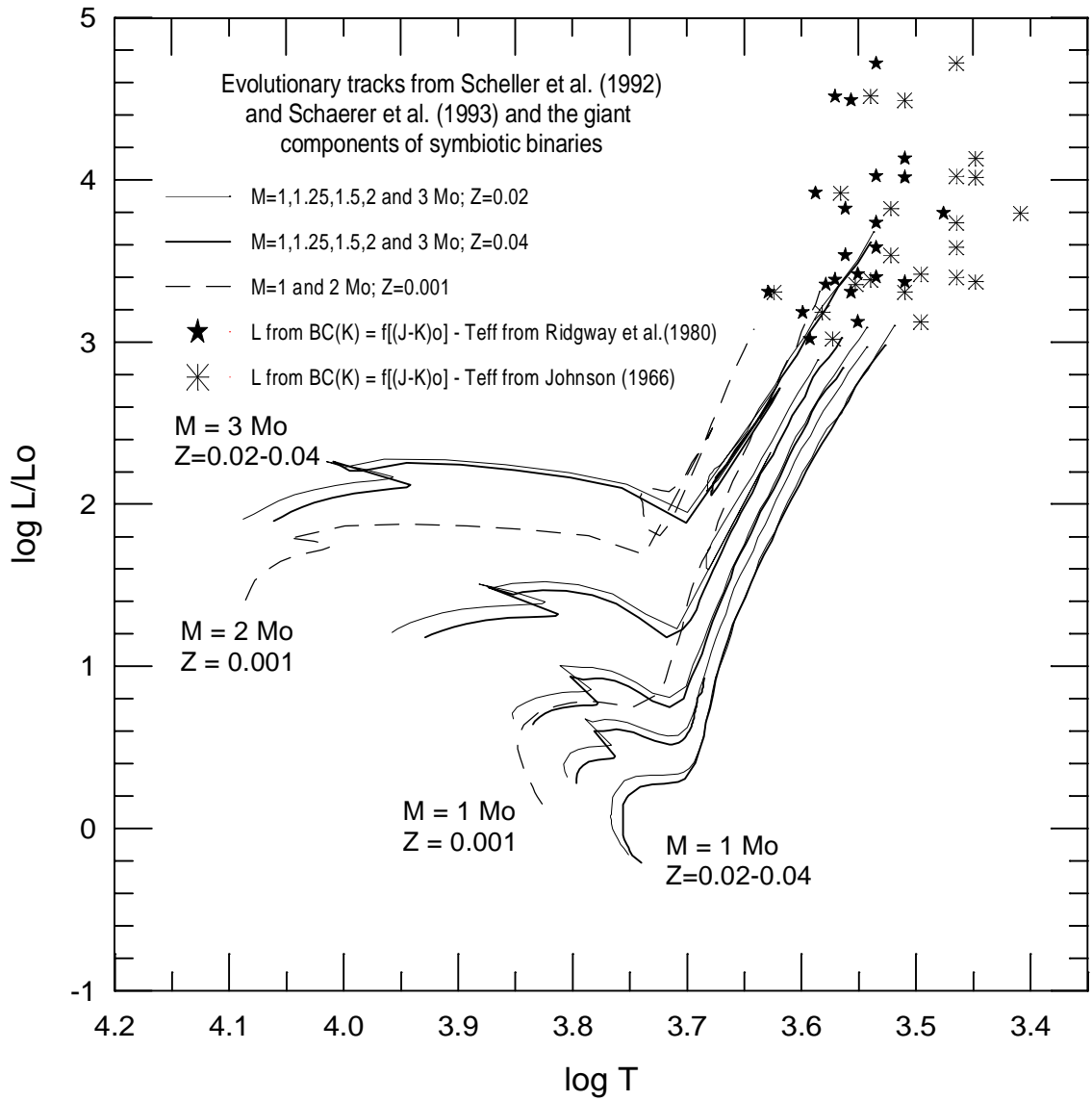

**Figure 6**

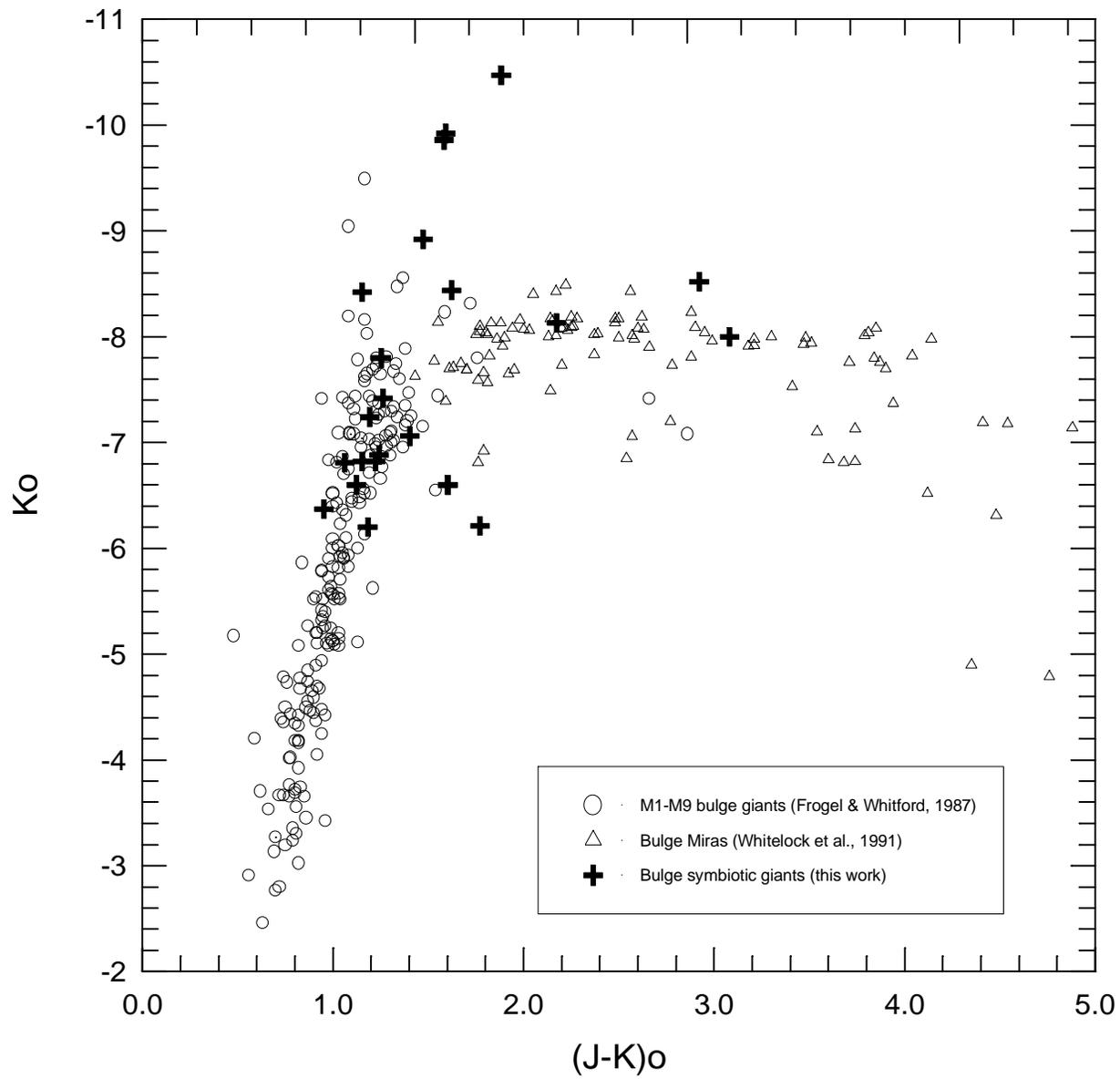

**Figure 7**